\documentclass[amstex,epsf,amssymb,usenatbib]{mn2e}
\usepackage{epsf}
\usepackage{amsmath,amssymb}
\usepackage{epsfig}
\usepackage{psfrag}

\newcommand{\bfm}[1]{\mbox{\boldmath $#1$}}
\newcommand{\ie}[1]{{\it i.e.}\ }
\newcommand{\etal}[1]{{\it et al.}\ }
\newcommand{\divb}[1]{\mbox{${\bfm \nabla}\cdot {\bf B}=0$}\ }
\newcommand{\beq}{\begin{equation}}
\newcommand{\eeq}{\end{equation}}

\newcommand{\mb}[1]{\mbox{\tiny{#1}}}
\newcommand{\ct}[1]{\mbox{cos}^{#1}\theta}
\newcommand{\st}[1]{\mbox{sin}^{#1}\theta}
\newcommand{\half}{\frac{1}{2}}
\newcommand{\n}{\noindent}
\newcommand{\bec}{\begin{center}}
\newcommand{\eec}{\end{center}}

\newcommand{\tbr}{\bar{\tau}}

\newcommand{\p}[2]{\frac{{\partial} #1}{\partial {#2}}}

\title{An explicit scheme for multifluid magnetohydrodynamics}
\author[S.\ O'Sullivan and T.P.\ Downes]
{Stephen O'Sullivan$^{1}$ \thanks{E-mail: stephen.osullivan@ucd.ie (SOS);
turlough.downes@dcu.ie (TPD)} and Turlough P. Downes$^{2}$\\
$^{1}$UCD School of Mathematical Sciences, University College Dublin, Belfield, Dublin 4, Ireland\\
$^{2}$School of Mathematical Sciences, Dublin City University, Glasnevin, Dublin 9, Ireland
}
\begin{document}

\date{Accepted .....
      Received ..... ;
      in original form .....}

\pagerange{\pageref{firstpage}--\pageref{lastpage}}
\pubyear{2005}

\maketitle

\label{firstpage}

\begin{abstract}

When modeling astrophysical fluid flows, it is often appropriate to discard the canonical magnetohydrodynamic approximation thereby freeing the magnetic field to diffuse with respect to the bulk velocity field.  As a consequence, however, the induction equation can become problematic to solve via standard explicit techniques.  In particular, the Hall diffusion term admits fast-moving whistler waves which can impose a vanishing timestep limit.

Within an explicit differencing framework, a multifluid scheme for weakly ionised plasmas is presented which relies upon a new approach to integrating the induction equation efficiently.  The first component of this approach is a relatively unknown method 
of accelerating the integration of parabolic systems by enforcing stability over large compound timesteps rather than over each of the constituent substeps.  This method, 
Super Time Stepping, proves to be very effective in applying a part of the Hall term up to a known critical value.  The excess of the Hall term above this critical value is then included via a new scheme for pure Hall diffusion.
\end{abstract}

\begin{keywords}
MHD -- shockwaves -- methods:numerical -- ISM:clouds -- dust, extinction
\end{keywords}

\section{Introduction}
\label{introduction}

Dynamically important magnetic fields are commonplace in astrophysics.
In many cases, where these fields interact with fluids, researchers have 
assumed that the equations of ideal magnetohydrodynamics (MHD) are sufficient 
in modeling the evolution of the magnetic fields and the fluids with which they 
interact.  There are clear examples, however, where the assumptions 
underpinning the equations of ideal MHD are not valid.  In dense
molecular clouds, for example, the density of charged particles can be
much lower than that of the neutral species \citep[][hereafter~CR02]{cr02}.  Under these conditions, coupling between
the motions of the fluids and the magnetic field is not
perfect, and diffusive effects become significant.  Similarly, ideal MHD is not believed to be 
valid in accretion disks around young stellar objects \citep[][]{wardle04}.  
The latter point is particularly interesting given the importance attached to the 
interaction between accretion disks and magnetic fields in the 
launching of stellar jets and outflows (e.g.\ \citealt{shu94}; \citealt{fc96}; 
\citealt{ouyed97}; \citealt{lery99}; \citealt{ferreira04}).  When modeling systems such as these therefore, a full multifluid treatment permitting relative motions between different component species should be adopted.

Many authors (\citealt{toth94}; \citealt{sm97}; \citealt{stone97}; 
\citealt{cff98}) have suggested schemes for numerically integrating the 
multifluid equations in the limit of pure ambipolar diffusion.
In this regime the charged species are firmly tied to the magnetic
field lines as they diffuse through the neutral gas.  The problem becomes more technically challenging, 
however, when charged species may be loosely attached to the field lines and Hall diffusion can become important.  Notably, it is thought that Hall diffusion may play an important role in environments such as the surfaces of
neutron stars (\citealt{hollrud04}), protostellar disks 
\citep[][]{wardle04}, and dense molecular clouds (CR02).

In their numerical studies of molecular clouds, CR02 assumed that the ionisation fraction is low 
and that the inertia of the charged particles may be neglected.  They were then
able to integrate the governing equations for a multifluid problem
including the presence of several species of charge-carrying grain.  Separately, Sano 
\& Stone (\citeyear{ss02a}; \citeyear{ss02b}) performed multifluid calculations 
designed to examine the Hall effect in the context of the magnetorotational 
instability in accretion disks.  However, both of the schemes used by these authors are subject to a rather stringent stability criterion which requires 
that the timestep tends to zero as the Hall effect becomes large
\citep[hereafter~F03]{falle03}.  To circumvent this constraint F03 presents a scheme employing an 
implicit method of integrating the magnetic field equation.  This has the advantage of allowing 
timesteps up to the limit dictated by the hyperbolic components of the equations.  
However, since large scale multifluid simulations are of obvious interest, the inherent difficulty of parallelising implicit schemes becomes a serious disadvantage. 

In this work we present a fully explicit numerical scheme for solving the 
multifluid equations describing a weakly ionised plasma.  The 
usual stability restrictions are relaxed through
a combination of a technique known as Super Time Stepping (STS) \citep{alex} and a 
new method which we call the Hall Diffusion Scheme (HDS).  Crucially, since the scheme is explicit, it is straightforward to parallelise and to implement on top of an adaptive mesh refinement (AMR) engine.

In Section~\ref{multifluid_equations} the governing equations are described; Section~\ref{numerical_approach} contains a
detailed description and analysis of the numerical scheme; Section~\ref{numerical_tests} contains numerical tests
demonstrating the reliability of the scheme; and in
Section~\ref{conclusions} the relevance of this work is discussed.

\section{The multifluid equations}
\label{multifluid_equations}

%The physics of the governing equations is outlined briefly in the following section.

We assume a weakly ionised plasma such that the mass density is dominated by the neutral component of the gas.  Then, relative to the scale-length of the system, if particles of a given charged species have small mean free paths in the neutral gas, or small Larmor radii, their pressure and inertia may be neglected. 

For convenience it is assumed there is no mass transfer between species.  It is straightforward, however, to insert the necessary terms for a more general treatment to include mass transfer (for example, see F03 and CR02) if desired.  The equations governing the evolution of the multifluid system (CR02; F03) can then be written as

\beq
\frac{\partial \rho_i}{\partial t} + \frac{\partial}{\partial x} \left(\rho_i
	\bfm{q}_i\right)  = 0 ,
\label{mass}
\eeq

\beq
\frac{\partial \rho_1 \bfm{q}_1}{\partial t} 
+ \frac{\partial}{\partial x}\left( \rho_1 u_1 \bfm{q}_1 + p_1\bfm{I}\right)  =  \bfm{J}\times\bfm{B} ,
\label{neutral_mom}
\eeq

\beq
\frac{\partial e_1}{\partial t} + \frac{\partial}{\partial
	x}\left[u_1\left(e_1+p_1 +\half\rho_1 q_1^2\right)\right] =\bfm{J}\cdot\bfm{E}  +\sum_{i=1}^N H_i  ,
	\label{neutral_en} 
\eeq

\beq
\frac{\partial \bfm{B}}{\partial t} + \frac{\partial \bfm{M}}{\partial
	x}  =  \frac{\partial}{\partial x} \bfm{R} \frac{\partial
		\bfm{B}}{\partial x}  ,
\label{B_eqn} 
\eeq

\beq
\alpha_i \rho_i\left(\bfm{E} + \bfm{q}_i \times \bfm{B} \right) +
\rho_i \rho_1 K_{i\,1}(\bfm{q}_1-\bfm{q}_i) =  0 , 
\label{charged_mom}
\eeq

\beq
H_i +G_{i\,1} +\alpha_i \rho_i \bfm{q}_i \cdot \bfm{E} =0 ,
\label{charged_en}
\eeq

\beq
\frac{\partial B_x}{\partial x} = 0 ,
\label{divB_eqn} 
\eeq

\beq
\sum_{i=2}^N \alpha_i \rho_i  =  0 ,
 \label{charge_neutrality}
\eeq

\beq
\sum_{i=2}^N \alpha_i \rho_i\bfm{q}_i  =  \bfm{J} .
 \label{current}
\eeq

\noindent The subscripts denote the species, with a subscript of 1
indicating the neutral fluid.  The variables $\rho_i$, $\bfm{q}_i \equiv
(u_i, v_i, w_i)^{\rm T}$ and $p_i$ are the mass
density, velocity and pressure of species~$i$.  The identity matrix, current density and
magnetic flux density are represented by $\bfm{I}$, $\bfm{J}$, $\bfm{B}$ respectively.  $K_{i\,1}$ describes the collisional interaction between species~$i$ and the neutral fluid, $\alpha_i$ is
the charge-to-mass ratio for species~$i$, $G_{i\,1}$ is the energy
transfer rate from species~$i$ to the neutral fluid, $H_i$ is the energy source or
sink appropriate to species~$i$, $\bfm{R}$ is the resistivity matrix and 
$\bfm{M}$ is the hyperbolic flux of $\bfm{B}$.  See F03 and
CR02 for a more detailed description of these terms.  Note that in
general $K_{i\,1}$ and $G_{i\,1}$ may depend on the temperatures and
relative velocities of the interacting species.  Equations~\eqref{mass}~to~\eqref{charged_en} are the
equations governing the conservation of mass, neutral momentum, 
neutral energy, magnetic flux, charged momentum, and charged energy.  Equations~\eqref{divB_eqn}~to~\eqref{current} describe the divergence of
$\bfm{B}$, charge neutrality, and current respectively.

From Faraday's law in one dimension $\partial B_x/\partial t = 0$ so that 
the trivial $B_x$ component may be dropped from equation~\eqref{B_eqn}.  The 
hyperbolic flux is then

\begin{equation}
\bfm{M} = (u_1B_y-v_1B_x, u_1B_z-w_1B_x) 
\end{equation}
\noindent and the resistivity matrix is
\begin{eqnarray}
\bfm{R} = & \nonumber \\ 
& \hspace*{-0.4in} \left(\begin{array}{ll}
		(r_O-r_A) \frac{B_z^2}{B^2} + r_A & (r_A-r_O)\frac{B_yB_z}{B^2} +
		r_H \frac{B_x}{B} \\
		(r_A-r_O)\frac{B_yB_z}{B^2} - r_H \frac{B_x}{B} &
		(r_O-r_A)\frac{B_y^2}{B^2} + r_A \end{array}\right)
\end{eqnarray}
where $r_O$, $r_H$ and $r_A$ are the Ohmic, Hall and ambipolar
resistivities respectively and are defined by 
\begin{eqnarray}
r_O & = & \frac{1}{\sigma_O} , \\
r_H & = & \frac{\sigma_H}{\sigma_H^2 + \sigma_A^2} , \\
r_A & = & \frac{\sigma_A}{\sigma_H^2 + \sigma_A^2} ,
\end{eqnarray}
\noindent with conductivities
\begin{eqnarray}
\sigma_O & = & \sum_{i=2}^N \alpha_i \rho_i \beta_i , \\
\sigma_H & = & \frac{1}{B}\sum_{i=2}^N \frac{\alpha_i \rho_i}{1+\beta_i^2} ,\\
\sigma_A & = & \frac{1}{B}\sum_{i=2}^N \frac{\alpha_i \rho_i \beta_i}{1+\beta_i^2} ,
\end{eqnarray}
\noindent where the Hall parameter for species~$i$ is given by
\begin{equation}
\beta_i = \frac{\alpha_i B}{K_{1\,i}\rho_1} .
\end{equation}

\section{Numerical approach}
\label{numerical_approach}

\subsection{The gas equations}

Assuming a piecewise constant solution at time $t^n$ on a uniform mesh
of spacing $h$, the solution at a later time $t^{n+1}=t^n+\tau$ is
sought.  The state in cell $j$ represents the volume average over 
$(j-1/2)h\le x\le (j+1/2)h$.

It should first be noted that the charged particle pressures\footnote{It is actually the charged species' temperatures which 
are derived as their pressures are not explicitly necessary under the 
assumptions made here.} and
velocities ($p_i^{n+1}$ and $\bfm{q}_i^{n+1}$ for~$i>1$) can be obtained
algebraically through equations~\eqref{charged_mom}~and~\eqref{charged_en}.  This procedure is described in Appendix~\ref{charged_vel_app}.

To obtain the full solution at time $t^{n+1}$, finite volume methods are 
applied to equations~\eqref{mass}~to~\eqref{charge_neutrality}.  The time 
integration is multiplicatively operator split into five operations, with 
each carried out to second order accuracy in space and time.  The order is 
permuted over successive timesteps such that second order temporal accuracy 
is maintained over the full step \citep{strang}.  In the following the five 
necessary operations for finite volume integration are described.

\begin{enumerate}
\item Equations~\eqref{mass}~to~\eqref{neutral_en} (with~$i=1$ for the mass equation) form a system of equations for the neutral gas.  Working in terms of the primitive variables $\bfm{P}=(\rho_1,\bfm{q}_1,\,p_1)^T$, fluxes are evaluated from a piecewise constant solution 
$\bfm{P}^{n}$ via a hydrodynamic Riemann solver.  A time centred
solution, $\bfm{P}^{n+1/2}$, obtained from these fluxes is then reconstructed 
to a second order piecewise-linear solution, $\bar{\bfm{P}}^{n+1/2}$, using 
Van Albada nonlinear averaging for the gradients.  Fluxes may then be derived 
from $\bar{\bfm{P}}^{n+1/2}$ which are second order accurate in space and 
time (for further details see, for example, ~\citealt{falle91}).  These
fluxes are then applied to the conserved variables.

\item The source terms on the right hand sides of equations~\eqref{neutral_mom}~and~\eqref{neutral_en} are applied.

\item The charged particle mass fluxes are applied using equation~\eqref{mass} with~$i>1$ in a second order upwind procedure similar to that used for the neutral gas.

\item The hyperbolic flux on the left hand side of equation~\eqref{B_eqn} is 
applied via a centred approximation

\beq
\bfm{M}_{j+1/2}=\half\left(\bfm{M}_{j+1}+\bfm{M}_{j}\right) .
\eeq

This has the disadvantage of not coupling the bulk fluid to the magnetic 
field through a Riemann problem, however, it is necessary in order that 
purely hydrodynamic subshocks may be properly captured.  As remarked by
F03, as long as the magnetic field appears continuous on the 
grid, as should be the case with finite resistivities, this is perfectly 
acceptable.

\item The resistive term on the right hand side of equation~\eqref{B_eqn} is applied.  Discussion of this procedure is deferred to the following section since it is of special 
interest.

\end{enumerate}

\subsection{Magnetic diffusion}

Splitting the hyperbolic flux term $\partial \bfm{M}/\partial x$ from the induction equation~\eqref{B_eqn} and linearising yields

\beq
\frac{\partial \bfm{B}}{\partial t} =  \bfm{R} \frac{\partial^2 \bfm{B}}{\partial x^2} . 
\label{Bdiff_eqn} 
\eeq

Note that the linearised form is assumed for convenience in the following analysis and in practise generalised discretisations of the nonlinear diffusion term are used. 

\subsubsection{Standard discretisation}

The usual explicit discretisation for a diffusion term applied to equation~\eqref{Bdiff_eqn} yields

\beq
\bfm{B}^{n+1}_j = \bfm{B}^{n}_{j} +\frac{\tau}{h^2}\bfm{R}^n_j(\bfm{B}^{n}_{j+1}-2\bfm{B}^{n}_{j}+\bfm{B}^{n}_{j-1}) 
\label{std}
\eeq

Assuming $r_O$ to be negligible, the relative importance of the remaining resistivities can be parameterised by \mbox{$\eta\equiv r_A/|r_H|$}.  F03 showed the above scheme has an amplification matrix with eigenvalues which are real when $\eta \ge \eta^*$ and complex otherwise.  The transition point $\eta^*$ is given by

\beq
\eta^*=2|\ct{}|/\st{2}
\eeq

\n where $\theta$ is the pitch angle of the field
with respect to the $x$-axis.  In the real regime, the stability limit on the 
timestep is 

\beq
\tbr^{\mb{R}}=\frac{2\sqrt{1+\eta^2}}{\eta(1+\ct{2})+2|\ct{}|\sqrt{(\eta/\eta^*)^2-1}}
\label{tauR}
\eeq

\n where $\tbr\equiv\tau/\tau^{\perp}$ and $\tau^{\perp}$ is the characteristic cell crossing time for diffusion perpendicular to the magnetic field given by

\beq
\tau^{\perp}=\frac{h^2}{2|r_H|\sqrt{1+\eta^2}} .
\eeq

\n However, below the transition point the stability limit becomes

\beq
\tbr^{\mb{C}}=\frac{1+\ct{2}}{2\ct{2}}\frac{\eta}{\sqrt{1+\eta^2}} .
\label{tauC}
\eeq

\n In either case the stable timestep limit goes as $h^2$ since this is an explicit
discretisation of a diffusion equation, however, a potentially more severe constraint is that while this 
limit increases as $\eta\to\eta^*$ in the real regime, it rapidly drops 
to zero as $\eta\to 0$ in the complex regime. 

\subsubsection{Numerical Strategy}

Our strategy is to split $r_H$ into two parts such that

\beq
r_H=r_H^a+r_H^b
\eeq

\noindent where $r_H^a\equiv\frac{\eta}{\eta^*}r_H$ is the maximum allowable
Hall resistivity in the real regime and $r_H^b$ is the excess.  The
induction equation is then integrated in two parts using a technique to accelerate the timestepping for the standard discretisation with Hall resistivity $r_H^a$.  The excess Hall resistivity $r_H^b$ is then applied using a different discretisation with suitable stability properties.

\subsubsection{Super Time Stepping}

STS is a technique which can be used to accelerate explicit schemes for
parabolic problems.  Essentially a Runge-Kutta-Chebyshev method, it has been known for some time (see \citealt{alex}), although it remains 
relatively unknown in computational astrophysics. 

A superstep $\tau^{\rm STS}$ is a composite timestep built up from a series 
of $N_{\rm STS}$ substeps such that

\beq
\tau^{\rm STS}=\sum^{N_{\rm STS}}_{j=1} d\tau_j .
\eeq

\noindent Judicious choice of the $d\tau_j$ yields stability for the 
superstep while the normal stability restrictions on the individual 
substeps are relaxed.  Exploiting the properties of Chebyshev polynomials provides a set of optimal values for the substeps given by
\beq
d\tau_j = \tau^X\left[ (-1+\nu)\cos\left(\frac{2j-1}{N_{\rm STS}}\frac{\pi}{2}\right) +1 +\nu\right]^{-1}
\eeq
\noindent where $\tau^X$ is the normal explicit timestep limit and $\nu$ is a
damping factor.  Note that $\tau^{\rm STS} \to N_{\rm STS}^2\tau^X$ as 
$\nu \to 0$.  The method is unstable in the limit $\nu=0$.  For a more 
detailed discussion, see \cite{alex} and references therein.

In order to apply STS to second order in time Richardson extrapolation in used.

\subsubsection{Hall Diffusion Scheme}

Having advanced the induction equation with a Hall resistivity
$r_H^a$, it is necessary to find an efficient scheme to impose the excess 
Hall diffusion $r_H^b$.  Since multiplicative operator splitting yields
a composite scheme with an amplification factor equal to the product of
the amplification factors of the basis schemes this task can be reduced to 
one of finding a scheme for pure Hall diffusion.

The key observation to make is that $\bfm{R}$ has zero entries on the 
diagonal when pure Hall diffusion is being considered.  With this in
mind, equation~\eqref{std} may be used to advance one component of 
the magnetic field explicitly, followed by an implicit-like discretisation of 
the alternate component.  We call this the Hall Diffusion Scheme (HDS) as we 
are not aware of an instance of this approach elsewhere in the literature.  
Hence the discretisation of equation~\eqref{std} for the pure Hall excess 
$r_H^b$ becomes

\beq
{B}^{n+1}_{y\,j} = {B}^{n}_{y\,j} +\frac{\tau}{h^2}{ d_H^b ({B}^{n}_{z\,j+1}-2{B}^{n}_{z\,j}+{B}^{n}_{z\,j-1})} 
\label{HDS1}
\eeq

\noindent followed by

\beq
{B}^{n+1}_{z\,j} = {B}^{n}_{z\,j} -\frac{\tau}{h^2}{ d_H^b ({B}^{n+1}_{y\,j+1}-2{B}^{n+1}_{y\,j}+{B}^{n+1}_{y\,j-1})} 
\label{HDS2}
\eeq
\n where the cosine term is absorbed by defining $d_H^b=r_H^b\ct{}$.  It seems to make little difference which component is advanced first. 

For clarity of notation the superscript $b$ is dropped from the following 
analysis of the stability properties of the scheme.  The resistance matrix 
for pure Hall diffusion is
\beq
\bfm{R}=
 \left( \begin{array}{cc}
0 & d_H  \\
-d_H & 0  
\end{array} \right). 
\eeq
\n Assuming a numerical wave of the form
\beq
\bfm{B}^n_j=\bfm{B}^n \mbox{e}^{i\omega j}
\eeq
\noindent in equations~\eqref{HDS1}~and~\eqref{HDS2} yields an amplification 
matrix
\beq
\bfm{A}=
 \left( \begin{array}{cc}
1 & -\hat{d}_H  \\
\hat{d}_H & 1-\hat{d}_H^2  
\end{array} \right)
\eeq 
\n where $\hat{d}_H\equiv\xi d_H$ and
\beq
\xi = \frac{2\tau(1-\mbox{cos}\omega)}{h^2}  .
\eeq
\n The eigenvalues of $\bfm{A}$ are given by
\beq
\lambda=1-\half\hat{d}_H^2 \pm i \half\hat{d}_H \sqrt{4-\hat{d}_H^2}
\eeq

\n and hence HDS is neutrally stable for $|\hat{d}_H|\le 2$.  Taking the most 
restrictive case of $\omega=\pi$ gives a stable timestep limit of
\beq
\tbr^{\mb{HDS}}=\frac{\sqrt{1+\eta^2}}{|\ct{}|(1-\eta/\eta^*)}  .
\label{tauH}
\eeq
\n Note that $\tbr^{\mb{HDS}} \to 1/|\ct{}|$ as $\eta\to 0$ in contrast to the 
standard discretisation for which $\tbr^{\mb{C}} \to 0$.  

The extension of the HDS to more than one dimension is straightforward although we defer a detailed discussion to a later paper. For an outline of the scheme in three dimensions the reader is referred to Appendix~\ref{3dhds}.

In practice, ordinary (unaccelerated) subcycling of HDS, using $N_{\rm HDS}$ subcycles, is applied in conjunction with STS.  This compound scheme (referred to as ``STS/HDS'' hereafter) usually allows the timestep limit imposed by the hyperbolic terms to be reached efficiently (see Subsection~\ref{comp_tim}).

\section{Numerical tests}
\label{numerical_tests}

Following F03, the dynamic algorithm described here is tested against 
solutions of the steady isothermal multifluid equations.  These steady 
state equations are solved using an independent code, the details of which 
are outlined in Appendix~\ref{steady_app}.  The conditions for each of the 
tests are given in Table~\ref{test_conditions}.

\begin{table*}
\begin{tabular}{llllll} \hline
Case A & & & & & \\
Right State & $\rho_1=1$ & $\bfm{q}_1 = (-1.751,0,0)$ & $\bfm{B} =
(1,0.6,0)$ & $\rho_2=5\times10^{-8}$ & $\rho_3 = 1\times10^{-3}$ \\
Left State & $\rho_1=1.7942$ & $\bfm{q}_1 = (-0.9759,-0.6561,0)$ & $\bfm{B} =
(1,1.74885,0)$ & $\rho_2=8.9712\times10^{-8}$ & 
$\rho_3 = 1.7942\times10^{-3}$ \\
 & $\alpha_2=-2\times10^{12}$ & $\alpha_3 = 1\times 10^8$ & $K_{2\,1} = 4
\times 10^5$ & $K_{3\,1} = 2 \times 10^4$ & $a=0.1$ \\
 & $\nu = 0.05$ & $N_{\rm STS} = 5$ & $N_{\rm HDS} = 0$ & & \\
Case B & & & & & \\
Right State & As case A & & & & \\
Left State & As case A & & & & \\
 & $\alpha_2=-2\times10^{9}$ & $\alpha_3 = 1\times 10^5$ & $K_{2\,1} = 4
\times 10^2$ & $K_{3\,1} = 2.5 \times 10^6$ & $a=0.1$ \\
 & $\nu = 0$ & $N_{\rm STS} = 1$ & $N_{\rm HDS} = 8$ & & \\
Case C & & & & & \\
Right State & $\rho_1=1$ & $\bfm{q}_1 = (-6.7202,0,0)$ & $\bfm{B} =
(1,0.6,0)$ & $\rho_2=5\times10^{-8}$ & $\rho_3 = 1\times10^{-3}$ \\
Left State & $\rho_1=10.421$ & $\bfm{q}_1 = (-0.6449,-1.0934,0)$ & $\bfm{B} =
(1,7.9481,0)$ & $\rho_2=5.2104\times10^{-7}$ & 
$\rho_3 = 1.0421\times10^{-2}$ \\
 & $\alpha_2=-2\times10^{12}$ & $\alpha_3 = 1\times 10^8$ & $K_{2\,1} = 4
\times 10^5$ & $K_{3\,1} = 2 \times 10^4$ & $a=1$ \\ 
 & $\nu = 0.05$ & $N_{\rm STS} = 15$ & $N_{\rm HDS} = 0$ & & \\ \hline
\end{tabular}
\caption{Test calculation parameters.}
\label{test_conditions}
\end{table*}

\subsection{Case A: Ambipolar Dominated}

\begin{figure}
  \bec
  \leavevmode
  \psfrag{p1}[Bl][Bl][1.2][-90]{\hspace*{-0.1in}$u_1$}
  \psfrag{p2}[Bl][Bl][1.2][  0]{\hspace*{-0.1in}$x$}
  \psfrag{p3}[Bl][Bl][1.2][-90]{\hspace*{-0.1in}$B_y$}
  \psfrag{p4}[Bl][Bl][1.2][  0]{\hspace*{-0.1in}$x$}
  \includegraphics[width=240pt]{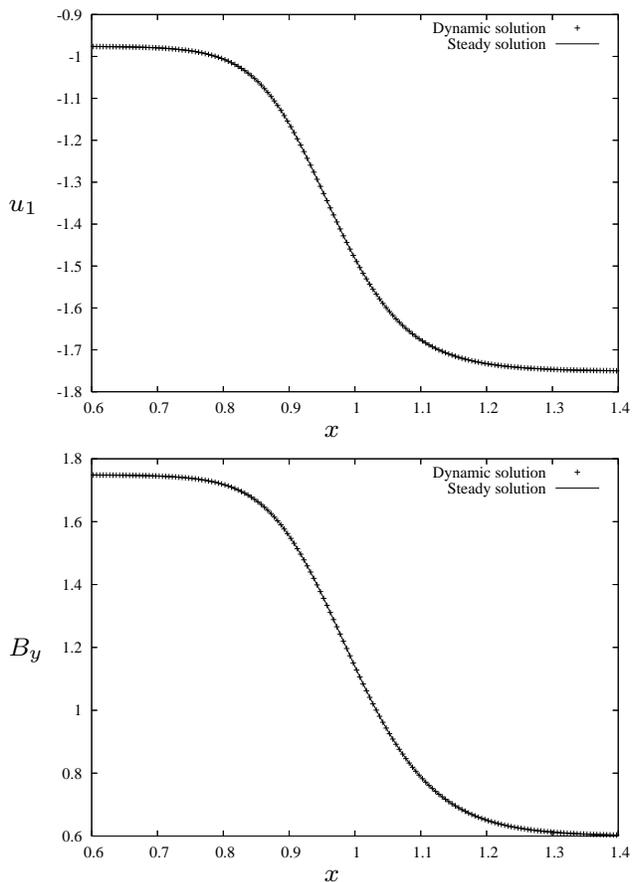}
  \caption{Neutral fluid $x$-velocity and $y$-component of magnetic field for case~A with $h = 5 \times 10^{-3}$.  The solution from the steady state equations, as a line, is overplotted with points from the dynamic code.}
  \label{fig_test_a}
  \eec
\end{figure}

In this test $r_{\rm O} = 2 \times 10^{-12}$, $r_{\rm H} = 1.16
\times 10^{-5}$ and $r_{\rm A} = 0.068$ giving $\eta = 5.86 \times 10^{3}$ and hence it can be expected that ambipolar diffusion will dominate the 
solution.  Fig.~\ref{fig_test_a} shows plots of the $x$ component of the neutral velocity, 
along with $B_y$ for both the dynamic and steady state solutions.  The 
calculation shown has $h = 5 \times 10^{-3}$.  It can be seen that the 
agreement between the two solutions is extremely good.

Since the algorithm is designed to be second order it is worthwhile
measuring the convergence rate of the dynamic solution against the solution from the steady state solver.  The comparison is made by minimising the L1 error 
norm, $e_1$, between a section of the dynamical solution and the steady state solution.  Working from the downstream side, the section $x_L\le x \le x_R$ is fixed about the point $x^*$ where the deviation from the downstream state first exceeds $50\%$ of the maximum variation in the solution.  Using $x_L=x^*-0.44$ and $x_R=x^*+0.56$ yields $e_1=3.90\times 10^{-5}$ for $h = 5 \times 10^{-3}$,
and $e_1=1.56\times 10^{-4}$ for $h = 1 \times 10^{-2}$.  This
gives $e_1 \propto h^{2.0}$ -- showing second order convergence 
as expected.

\subsection{Case B: Hall Dominated}

\begin{figure}
  \bec
  \leavevmode
  \psfrag{p1}[Bl][Bl][1.2][-90]{\hspace*{-0.1in}$u_{\mbox{\tiny{$1$}}}$}
  \psfrag{p2}[Bl][Bl][1.2][  0]{\hspace*{-0.1in}$x$}
  \psfrag{p3}[Bl][Bl][1.2][-90]{\hspace*{-0.1in}$B_y$}
  \psfrag{p4}[Bl][Bl][1.2][  0]{\hspace*{-0.1in}$x$}
  \includegraphics[width=240pt]{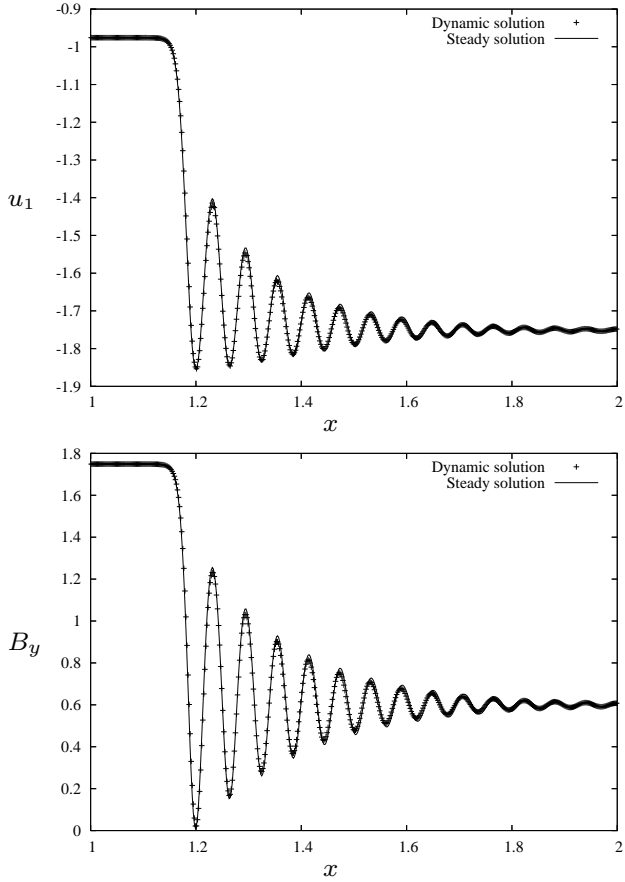}
  \caption{Neutral fluid $x$-velocity and $y$-component of magnetic field for case~B with $h = 2 \times 10^{-3}$.  The solution from the steady state equations, as a line, is overplotted with points from the dynamic code.}
  \label{fig_test_b}
  \eec
\end{figure}

The Hall term dominates in this test, requiring the Hall diffusion to be split and applied in part via HDS.  The parameters are $r_{\rm O}=2\times10^{-9}$, $r_{\rm H}=0.0116$,
$r_{\rm A}=5.44\times10^{-4}$ with $\eta = 0.0046 \ll 1$.\footnote{If the Hall diffusion is increased much further, it appears that the approximation of negligible charged particle inertia breaks down.}  
Fig.~\ref{fig_test_b} shows the results of the calculations for 
the test with $h = 2 \times 10^{-3}$.  For standard explicit codes 
the conditions lead to prohibitive restrictions on the timestep.  However, the use of HDS allows us to 
maintain a timestep close to the Courant limit imposed by the hyperbolic terms throughout the 
calculations.

As with case~A, the dynamic solution is tested to ensure it has the correct 
second order convergence characteristics.  With $x_L=x^*-0.15$ and $x_R=x^*+0.95$, we find $e_1 = 4.95\times10^{-3}$ for $h = 2\times10^{-3}$ and $e_1 = 1.15\times10^{-3}$ for 
$h = 1 \times 10^{-3}$ , giving 
$e_1 \propto h^{2.1}$.  Again, this is close to the second order convergence rate expected.

\subsection{Case C: Neutral subshock}

This test is similar to case~A, but with a higher soundspeed and upstream 
fast Mach number.  As a result, a subshock develops in the neutral flow because the 
interactions between the charged particles and the neutrals are not strong 
enough to completely smooth out the strong initial discontinuity in the 
neutral flow.  The ability of the algorithm described to deal with discontinuities in the solution is therefore tested.

\begin{figure}
  \bec
  \leavevmode
  \psfrag{p1}[Bl][Bl][1.2][-90]{\hspace*{-0.1in}$u_{\mbox{\tiny{$1$}}}$}
  \psfrag{p2}[Bl][Bl][1.2][  0]{\hspace*{-0.1in}$x$}
  \psfrag{p3}[Bl][Bl][1.2][-90]{\hspace*{-0.1in}$B_y$}
  \psfrag{p4}[Bl][Bl][1.2][  0]{\hspace*{-0.1in}$x$}
  \includegraphics[width=240pt]{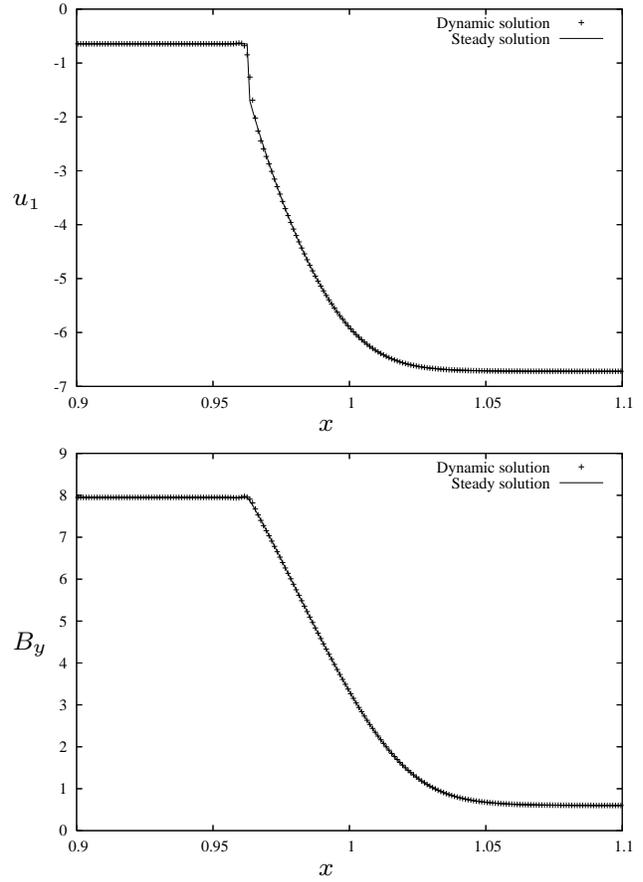}
  \caption{Neutral fluid $x$-velocity and $y$-component of magnetic field for case~C with $h = 1 \times 10^{-3}$.  The solution from the steady state equations, as a line, is overplotted with points from the dynamic code.}
  \label{fig_test_c}
  \eec
\end{figure}

\begin{figure}
  \bec
  \leavevmode
  \psfrag{p1}[Bl][Bl][1.2][-90]{\hspace*{-0.1in}$u_{\mbox{\tiny{$2$}}}$ }
  \psfrag{p2}[Bl][Bl][1.2][  0]{\hspace*{-0.1in}$x$}
  \psfrag{p3}[Bl][Bl][1.2][-90]{\hspace*{-0.1in}$B_y$}
  \psfrag{p4}[Bl][Bl][1.2][  0]{\hspace*{-0.1in}$x$}
  \includegraphics[width=240pt]{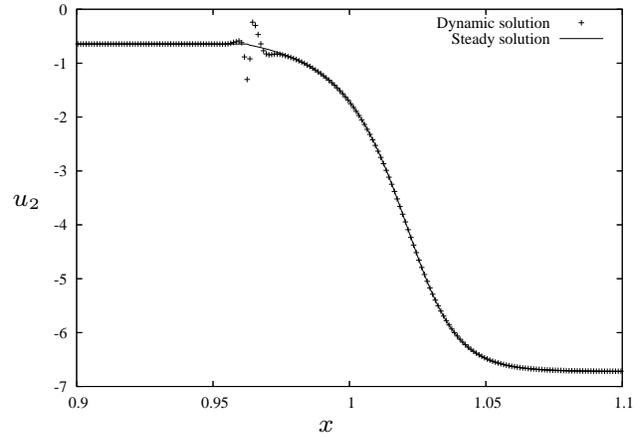}
  \caption{Negatively charged fluid $x$-velocity for case~C with $h = 1 \times 10^{-3}$.  The solution from the steady state equations, as a line, is overplotted with points from the dynamic code.}
  \label{fig_test_c_2}
  \eec
\end{figure}

Fig.~\ref{fig_test_c} shows the results of the calculations for
$h = 1 \times 10^{-3}$.  The subshock in the neutral flow is
clearly visible as a discontinuity in $u_1$, while there is no
corresponding discontinuity in $B_y$.  Fig.~\ref{fig_test_c_2}
contains a plot of the $x$ component of the velocity of the negatively 
charged fluid.  As expected, there is no discontinuity in this variable,
but there are some oscillations at the point where the discontinuity in
the neutral flow occurs.  These errors are remarkably similar to those
encountered by F03 and do not affect the global solution.

It can be expected that, since there is a discontinuity in the solution of this
test and a MUSCL-type approach is used, the rate of convergence of the
dynamic solution will be close to first order, at least for resolutions high 
enough to discern the subshock in the solution.  In this test 
$x_L=x^*-0.13$ and $x_R=x^*+0.15$.  We find $e_1 = 3.41\times10^{-2}$ for $h = 5\times10^{-3}$ and $e_1 = 5.25\times10^{-3}$ for $h = 1\times10^{-3}$ yielding 
$e_1 \propto h^{1.16}$ -- close to the first order expected, although 
clearly the error from around the subshock is not completely dominating at 
this resolution.  At $h = 5 \times 10^{-4}$ we find $e_1 = 2.73\times10^{-3}$ 
giving $e_1 \propto h^{0.94}$ with respect to the error at $h = 1\times10^{-3}$.  We suspect that the deviation from first order is due to a discontinuity in the electric field at the subshock causing an error in the charged velocities since smoothing the solution with some artificial viscosity is found to improve the convergence.

\subsection{Comparative timings}
\label{comp_tim}

In this section comparison is made between the performances of standard explicit, STS/HDS, and implicit (Crank-Nicolson) discretisations of the induction equation.  The different methods are applied in otherwise identical codes to the high resolution trials of the preceding test cases.  Since the neutral gas equations are treated explicitly in all cases, the corresponding Courant condition on the integration of the hyperbolic terms imposes a hard limit on the timestep.

\begin{table}
\begin{tabular}{llll} \hline 
 & Case A & Case B & Case C \\ \hline
STS/HDS  & 1.9 &   14.8   & 1.9 \\
Implicit & 1.9 &   23.3   & 2.7 \\ \hline
\end{tabular}
\caption{\label{timings} The speed-up factors in CPU time usage achieved via the implicit and STS/HDS discretisations of the induction equation relative to the standard explicit discretisation.}
\label{tab_tim}
\end{table}

As a benchmark, we use the standard explicit discretisation subcycled to the same degree as the STS/HDS method. The speed-up factors of the STS/HDS and implicit methods in terms of CPU time usage are presented in Table~\ref{tab_tim}.  Clearly, either technique offers a significant improvement in efficiency and both achieve timesteps close to the limit introduced by the hyperbolic terms.  The implicit method is slightly faster for case~C due to the high degree of subcycling used for the STS and significantly so for case~B because of the very large Hall term. Otherwise, the  STS/HDS and implicit methods yield similar speed-up factors indicating that overall efficiency is dominated by the other parts of the schemes. It should be emphasised that these are steady state problems which suit implicit methods particularly well and for non-steady state problems accuracy constraints may reduce the efficiency of implicit schemes.

\section{Conclusions}
\label{conclusions}

A new explicit scheme for integrating the multifluid equations in the limit of low ionisation has been presented.  The usual explicit stability limit imposed by the induction equation is relaxed by means of the Super Time Stepping algorithm applied for a portion of the Hall diffusion up to a critical limiting value.

Beyond this limiting value the standard explicit discretisation becomes
subject to a stability constraint requiring that the timestep vanish as the
Hall diffusion becomes large.  In order to circumvent this constraint, the 
excess Hall diffusion above the critical value is split off and applied via a 
new method which we have called the Hall Diffusion Scheme.

It has been demonstrated that, for the case of an isothermal flow, the
algorithm is accurate and converges to second order when
the solution is smooth and to first order when the solution contains a
discontinuity.  The extension of this scheme to non-isothermal flow does not 
present any obvious difficulties, although a modification of the 
discretisation used for the magnetic flux evolution is necessary.

Since all discretisations used in the scheme presented here are explicit, it is a straightforward matter to implement in a multidimensional parallelised codes using adaptive mesh refinement.  This is a crucial advantage for large-scale simulations of astrophysical systems in which multifluid effects are thought to be important such as dense molecular clouds and protostellar accretion disks.

\section*{Acknowledgements}          

This work was partly funded by the CosmoGrid project, funded under the
Programme for Research in Third Level Institutions (PRTLI) administered by the 
Irish Higher Education Authority under the National Development Plan and with 
partial support from the European Regional Development Fund.

The authors are grateful to the School of Cosmic Physics at the Dublin
Institute for Advances Studies for facilitating this collaboration and
to the referee Sam Falle for valuable comments.

\appendix

\section{Charged velocities}
\label{charged_vel_app}

For this work the collisional coefficients $K_{i\,1}$ are assumed to be 
independent of velocities and temperatures.  The following derivation (S.A.E.G. Falle, private communication) is included for completeness.

Transforming to the frame co-moving with the neutral gas, equation~\eqref{charged_mom} can be written as

\beq
\bfm{q}'_i \times \bfm{B} -\kappa_i\bfm{q}'_i = -\bfm{E}' -\bfm{q}_1\times\bfm{B}
\eeq

\noindent where $\kappa_i\equiv\rho_1 K_{i\,1}/\alpha_i$ and $\bfm{E}'=\bfm{E}+\bfm{q}_1\times\bfm{B}$.

Then choosing~$i=2$ as a reference species, the general solutions for the remaining charged species' velocities are given by

\beq
\bfm{q}'_i=\bfm{A}^{-1}_i\bfm{A}_2\bfm{q}'_2
\eeq

\noindent where

\beq
\bfm{A}_i = \left(\begin{array}{ccc}

-\kappa_i & B_z & -B_y \\
-B_z & -\kappa_i & B_x \\
B_y & -B_x & -\kappa_i 

\end{array}\right) .
\eeq

To derive the charged velocities, all that remains is for the reference velocity to be evaluated, this can be done by using equation~\eqref{current} and Amp\`{e}re's law to give

\beq
\bfm{q}'_2=\left[  \bfm{I} -\left(\sum_{i=3}^N \frac{\alpha_i \rho_i}{\alpha_2\rho_2}\bfm{A}^{-1}_i\right)\bfm{A}_2           \right]^{-1}  \frac{\nabla\times\bfm{B}}{\alpha_2\rho_2} .
\eeq

If the collisional coefficients are in fact dependent on the velocities of the charged species, this procedure can be carried out iteratively using the values from the previous timestep as a starting point.

Should the collisional coefficients depend on the temperatures of the 
charged species, some additional calculation is necessary before the next 
iteration: using equation~\eqref{charged_en} and inserting the specific form of 
the function $G_{i\,1}$, $N-1$ equations are obtained which may be solved readily for the $N-1$ charged 
temperatures. 

Finally, it is worth noting that superior results are obtained by 
interpolating the primitive quantities to the cell edges before calculating 
the charged velocities rather than by calculating the velocities at the cell 
centres and interpolating from these to the edges.

\section{HDS in three dimensions}
\label{3dhds}

%In the following derivation we make a temprary switch in notation for clarity and refer the vector components with numeral indices such that $x_1\equiv x$,  $x_2\equiv y$, and  $x_3\equiv z$.

Equation~\eqref{charged_mom} can be used in conjunction with with equation~\eqref{current} to write the electric field for pure Hall diffusion as

\beq
\bfm{E}=r_H\frac{\bfm{J}\times\bfm{B}}{B}
\eeq

\n Then, using Faraday's law, we can write 

\beq
\p{\bfm{B}}{t}=-r_H\nabla\times(\bfm{J}\times\bfm{b})
\eeq

\n where $\bfm{b}\equiv\bfm{B}/B$.

\n This equation can be expanded out and linearised to give

\beq
\p{\bfm{B}}{t}=\bfm{G}\bfm{B}
\eeq

\n where, using $\bfm{J}=\nabla\times\bfm{B}$, the matrix $\bfm{G}$ is given by

\beq
\bfm{G} = -r_H(\bfm{b}\cdot\nabla)\nabla\times .
\eeq

\vspace*{0.2in}

\n Hence \bfm{G} is antisymmetric and we can write the generalised HDS scheme as

\bec
\begin{eqnarray}
B_x^{n+1} &=& B_x^n + \tau(G_{x\,y}^n B_y^n +  G_{x\,z}^n B_z^n) ,
\label{HDSgen1} \\
B_y^{n+1} &=& B_y^n + \tau(G_{y\,z}^n B_z^n +  G_{y\,x}^n B_x^{n+1}) ,
\label{HDSgen2} \\
B_z^{n+1} &=& B_z^n + \tau(G_{z\,x}^n B_x^{n+1} +  G_{z\,y}^n B_y^{n+1}) ,
\label{HDSgen3}
\end{eqnarray}
\eec

\n where $\bfm{G}^n$ is the discretised form of the operator $\bfm{G}$ at time level $n$. 

The generalised HDS scheme in three dimensions is analogous in construction to the one-dimensional case in that equation~\eqref{HDSgen1} is an explicit first step and equation~\eqref{HDSgen3} is an implicit-like final step. Additionally, we now have an intermediate step of mixed explicit/implicit character. Numerical tests indicate that the method retains its favourable stability properties in three dimensions.

\section{Steady-state solver}
\label{steady_app}

Assuming an isothermal flow, as is the case for the tests presented in this 
work, setting all derivatives with respect to time to zero in the multifluid 
equations gives us 
\begin{eqnarray}
\rho_i u_i = Q_i = \rm{constant} , \\
\rho_1 u_1^2+a^2\rho_1 + \frac{B^2}{2} = P_x = \rm{constant} , \\
\rho_1 u_1 v_1 - B_x B_y = P_y = \rm{constant} , \\
\rho_1 u_1 w_1 - B_x B_z = P_z = \rm{constant} . 
\end{eqnarray}
In addition the reduced momentum equations for the charged species 
(equation~\eqref{charged_mom}) yield 3 equations for each charged species, 
and the charge neutrality condition is also used.  Finally, the
equation for $\bfm{B}$ yields
\begin{equation}
\bfm{M} - \bfm{M_{\rm R}} = \bfm{R}\frac{d\bfm{B}}{dx}
\end{equation}
with $\bfm{M}_{\rm R}$ ($ = \bfm{M}_{\rm L}$) being the flux in the
right (left) state.

For the cases considered here, with two charged species, the above
equations constitute one ordinary differential equation for $\bfm{B}$
and seven equations which, once $\bfm{B}$ is known at a given point in
space, can be used to solve for all the other variables.  The ODE for 
$\bfm{B}$ is solved using the Runge-Kutta method of order~4.

The initial conditions (at $x=0$) are a saddle point of the ODE for
$\bfm{B}$.  These conditions are perturbed slightly and the system then
evolves through phase space to a sink point.

\label{lastpage}

\end{document}